\documentclass[12pt]{article}
\usepackage{amsmath}
\usepackage{amssymb}
\usepackage{amstext}
\usepackage{amsfonts}
\usepackage{graphicx,epsfig}
\usepackage{slashbox}
\usepackage{indentfirst}
\begin{document}

\baselineskip 20pt

\title{Discussions on Stability of Diquarks }
\author{ Liang Tang and Xue-Qian Li\\[0.5cm]
{\small{School of Physics, Nankai University, 300071, Tianjin,
China}}}
\date{}
\maketitle

\begin{abstract}
Since the birth of the quark model, the diquark which is composed of
two quarks has been considered as a substantial structure of color
anti-triplet. This is not only a mathematical simplification for
dealing with baryons, but also provides a physical picture where the
diquark would behave as a whole object. It is natural to ask whether
such a structure is sufficiently stable against external
disturbance. The mass spectra of the ground states of the scalar and
axial-vector diquarks which are composed of two-light (L-L),
one-light-one-heavy (H-L) and two-heavy quarks (H-H) respectively
have been calculated in terms of the QCD sum rules. We suggest a
criterion as the quantitative standard for the stability of the
diquark. It is the gap between the masses of the diquark and
$\sqrt{s_0}$ where $s_0$ is the threshold of the excited states and
continuity, namely the larger the gap is, the more stable the
diquark would be. In this work, we calculate the masses of the type
H-H to complete the series of the spectra of the ground state
diquarks. However, as the criterion being taken, we find that all
the gaps for the various diquaks are within a small range,
especially the gap for the diquark with two heavy quarks which is
believed to be a stable structure, is slightly smaller than that for
other two types of diquarks, therefore we conclude that because of
the large theoretical uncertainty, we cannot use the numerical
results obtained with the QCD sum rules to assess the stability of
diquarks, but need to invoke other theoretical framework.

\noindent{PACS numbers: 11.55.Hx, 12.38.Lg}\\
\end{abstract}

\section{Introduction}
Right after the birth of the quark model, the diquark model was
proposed: two quarks constitute a color-anti-triplet which may be a
tightly bound state. In Gell-Mann's pioneer paper on the quark
model, he discussed the possibility of existence of free
diquarks\cite{GellMann:1964nj}. With the diquark picture, numerous
authors have studied the processes where baryons are involved
\cite{Majethiya:2008ia,Tong:1999qs, Ebert:2002ig,
He:2004px,Kiselev:2002iy} and their conclusions support the
existence of diquarks. Even in  the meson sector, some newly
observed resonances are considered to possess the exotic structures.
One possibility is that the mesons are tetraquarks composed of a
diquark and an anti-diquark, \cite{Maiani:2004vq,
Nielsen:2005ia,Ding:2006vk,Zhang:2006xp,Rehman:2011am}. In fact, it
is still in dispute that the diquark is a real spatially bound state
as a pseudo-particle or just a loosely bound state. Recently, the
authors of Refs.\cite{Kim:2011ut,Tang:2011fv} treat the diquark as
an explicit particle which is the essential ingredient inside
hadrons (baryons or exotic mesons). If these diquarks indeed exist
as stable particles, they should have certain and definite masses
and quantum numbers. Just as we discuss the regular hadrons, their
spectra not only possess the real part, but also the imaginary part
which corresponds to the stability of the diquark. Namely, the
lifetime of the diquark might be finite. The main scenario to
determine the diquark lifetime is that with an external disturbance,
the diquark might dissolve into two quarks. By our intuition, the
diquark composed of two light quarks might be easier to dissolve by
absorbing gluons. Generally, it is believed that the heavy diquarks
which are composed of two heavy quarks are more stable against
external disturbance.

In principle, a baryon which is composed of three valence quarks is
described by the Faddeev equation group composed of three coupled
differential equations\cite{Ishii:1995bu,Nicmorus:2008eh}. For the
equations, the three valence quarks are of the same weights. It is
possible that two of the three quarks would accidentally  constitute
a bound state, say, by quantum fluctuation. The diquark can be
treated as a sub-system and behaves as an independent particle. In
this picture the baryon possesses a diquark-quark structure and thus
the three-body system turns into a two-body one. Correspondingly,
the three Faddeev equations reduce to a single equation (no mater
relativistic or non-relativistic). Therefore the problem is greatly
simplified. It is noted on other side, that the sub-system is not
exactly a fundamental one, but possesses an inner structure. When it
interacts with gluons, a form factor which manifests the inner
structure should be introduced. Now, we are confronting a problem:
which two valence quarks of the three would tend to combine into a
bound state. In fact, any two quarks have a chance to combine via
strong interactions, but the rest valence quark would interact with
the individual quarks in the diquark and tend to tear it apart. Thus
the key point is if such sub-system is sufficiently stable against
the disturbance. When we deal with the baryons which are composed of
three light quarks, one-heavy-two-light quarks and
two-heavy-one-light quarks, we notice an obvious unequal structures.
We need to determine which type of diquark  i.e. the diquarks with
light-light (L-L), heavy-light (H-L) or heavy-heavy (H-H)
structures, is more stable. Thus we can more confidently reduce the
three Faddeev equations to one with the expected diquark subsystem.
For serving this aim, it is significant to investigate the stability
of diquarks.

First of all, it is important and interesting to investigate the
spectra of the diquarks  with various quark contents. According to
the masses of different flavors, one can categorize the diquarks
into three types: light-light diquark (L-L), light-heavy diquark
(H-L) and heavy-heavy diquark (H-H), where light quarks are $u, d,
s$ and heavy quarks are $c, b$.

The QCD is by all means a successful theory for the strong
interaction, but the non-perturbative QCD which dominates the low
energy phenomena is still not fully understood yet. Among the
theoretical methods for treating the non-perturbative QCD effects,
the QCD Sum Rules\cite{Shifman} is believed to be a powerful means
for evaluating the hadronic spectra and other properties of hadrons.
In Refs.\cite{Zhang:2006xp,Dosch:1988huJamin:1989hh}, the mass
spectra of the scalar light-light diquark states (L-L) were studied
with the QCD sum rules. Recently, with the same method, Wang studied
the light-heavy diquark states (H-L)\cite{Wang:2010sh}.

By the common sense, the diquark  composed of two heavy quarks (H-H)
may be a kinematically favorable sub-structure in baryon. For
studying the stability of the three different types (L-L, H-L and
H-H) diquarks, it would be crucial to discuss their properties in a
unique framework. In this work, we are going to carry out the job in
terms of the QCD sum rules. Then we will discuss the feasibility of
such a scheme by scanning the numerical results obtained in this
method.

In this paper, we calculate the masses of heavy-heavy diquark states
(H-H) with the QCD sum rules. Then combining the results presented
in the relevant works \cite{Dosch:1988huJamin:1989hh,Wang:2010sh},
we discuss the stability of the diquarks altogether.

Obviously it is crucial to set a reasonable and practical criterion
of the stability of the sub-structures $-$ diquarks.

In the scenario of the QCD sum rules, numerically there exists a
threshold  $s_0$ corresponding to a starting point  beyond which
higher excited and continuous states reside. This cutoff provides a
natural criterion which we may use to study the stability of the
diquark. Namely, according to our general knowledge on quantum
mechanics, the continuous spectra would correspond to the dissolved
state where the constituents of the supposed-bound state would be
set free. Thus we choose the gap between the ground state and the
corresponding threshold $\sqrt{s_0}$ as the criterion of stability
of the diquarks.

However, as well known that there is a 20\% theoretical uncertainty
in all the computations via the QCD sum rules, therefore, even
though such criterion may be indeed reasonable in principle, it is
still doubtful if the results obtained in terms of the QCD sum rules
can practically apply to reflect the diquark stability. The goal of
this work is to testify the reasonability of applying the supposed
criterion within the framework of the QCD sum rules.

In the last section, we will come back to discuss the feasiability
based on the numerical results we obtained in term of the QCD sum
rules.

The paper is organized as follows. After the introduction, in Sec.II we derive
the correlation function of the suitable
currents with proper quantum numbers in terms of the QCD sum rules.
In Sec. III, our numerical results and relevant figures are
presented. Section IV is devoted to a summary and concluding
remarks. The tedious analytical results are collected in the
appendices.

\section{Formalism}
For studying the scalar and axial-vector H-H diquark  with the QCD
sum rules, we write down the local color anti-triplet diquark
currents:
\begin{eqnarray}
J^i(x)&=&\epsilon^{ijk}Q^T_j(x)C\gamma^5Q_k(x)\;,\\
J^i_\mu(x)&=&\epsilon^{ijk}Q^T_j(x)C\gamma_\mu Q_k(x)\;,
\end{eqnarray}
where  $i, j, k$ are the color indexes, $Q=b,c$, and  $C$ is the
charge conjugation operator.

In order to perform the QCD sum rules, we define the
two-point correlation functions $\Pi(q)$ (for scalar diquark) and
$\Pi_{\mu\nu}(q)$ (for axial-vector diquark) as follows:
\begin{eqnarray}
\Pi(q)&=&i\int d^4x e^{iq\cdot x}\langle
0|T\Big{\{}J^i(x)J^{i\dagger}(0)\Big{\}}|0\rangle\;,\\
\Pi_{\mu\nu}(q)&=&i\int d^4x e^{iq\cdot x}\langle
0|T\Big{\{}J^i_\mu(x)J^{i\dagger}_\nu(0)\Big{\}}|0\rangle\;.
\end{eqnarray}

On the hadron side, after separating out the ground state
contribution from the pole terms, the correlation function is
expressed as a dispersion integral over the physical regime,
\begin{eqnarray}
\Pi(q)&=&\frac{\lambda_S^2}{M_S^{2}-q^2}+\frac{1}{\pi}\int_{s^0_S}^\infty
ds\frac{\rho_S^h(s)}{s-q^2}\;,\label{sds}\\
\Pi_{\mu\nu}(q)&=&\Big(-g_{\mu\nu}+\frac{q_\mu
q_\nu}{q^2}\Big)\bigg\{\frac{\lambda_A^2}{M_A^{2}-q^2}+\frac{1}{\pi}\int_{s^0_A
}^\infty ds\frac{\rho_A^h(s)}{s-q^2}\bigg\}\;,\label{sda}
\end{eqnarray}
where $M_t$ with subscript t being S or A for the scalar or
axial-vector respectively, is the mass of the ground state diquark,
$\rho^h_{t}(s)$ is the spectral density and represents the
contribution from the higher excited states and the continuum,
$s^0_t$ is the threshold for the excited states and continuum, the
pole residues $\lambda_t$ correspond to the diquark coupling
strength, is defined through\cite{Wang:2010sh}:
\begin{eqnarray}
\langle0|J^i(0)|S^j(q)\rangle&=&\lambda_S\delta^{ij}\;,\\
\langle0|J^i_\mu(0)|A^j(q)\rangle&=&\lambda_A\epsilon_\mu\delta^{ij}\;,
\end{eqnarray}
with $\epsilon_\mu$ being the polarization vector of the
axial-vector diquark.

On the quark side, the operator product expansion (OPE) is applied
to derive the correlation functions. Firstly, we can write down the
``full" propagator $S_Q^{ij}(x)$ of a massive quark, where the
vacuum condensates are clearly displayed\cite{Reinders:1984sr}.
\begin{eqnarray}
S_Q^{ij}(x)&=&\frac{i}{(2\pi)^4}\int d^4p e^{-ip\cdot x} \left\{
\frac{\delta_{ij}}{\!\not\!{p}-m_Q}
-\frac{g_s(t^k)^{ij}G^{\alpha\beta}_{k}}{4}\frac{\sigma_{\alpha\beta}(\!\not\!{p}
+m_Q)+(\!\not\!{p}+m_Q)\sigma_{\alpha\beta}}{(p^2-m_Q^2)^2}\right.\nonumber\\
&&\left.+\frac{\pi^2}{3} \langle \frac{\alpha_sGG}{\pi}\rangle
\delta_{ij}m_Q \frac{p^2+m_Q\!\not\!{p}}{(p^2-m_Q^2)^4}
+\cdots\right\}\;,
\end{eqnarray}
where
$\langle\frac{\alpha_sGG}{\pi}\rangle=\langle\frac{\alpha_sG_{\alpha
\beta}G^{\alpha\beta}}{\pi}\rangle$, then contracting the quark fields
in the correlation functions, we gain the
results:
\begin{eqnarray}
\Pi(q)&=&-i\epsilon^{ijk}\epsilon^{ij^\prime k^\prime}\int d^4x
e^{iq\cdot x}Tr\bigg\{\gamma_5S_Q^{jj^\prime}(x)\gamma_5CS_Q^{k
k^\prime T}(x)C\bigg\}\;,\\
\Pi_{\mu\nu}(q)&=&i\epsilon^{ijk}\epsilon^{ij^\prime k^\prime}\int
d^4xe^{iq\cdot x}Tr\bigg\{\gamma_\mu S_Q^{jj^\prime}(x)\gamma_\nu
CS_Q^{kk^\prime T}(x)C\bigg\}\;.
\end{eqnarray}

Then substituting the full $c$ and $b$ quark propagators into above
correlation functions and integrating over the variable $k$, we
obtain the correlation functions at the level of quark-gluon degrees
of freedom. Simply, the correlation function $\Pi_t(q^2)$ (t=S or A)
is written as:
\begin{eqnarray}
\Pi_t(q)=\Pi_t^{\text{pert}}(q)+ \Pi_t^{\text{cond,4}}(q)\;,
\end{eqnarray}
where the superscripts ``pert", and``cond"  refer to the
contribution from the perturbative QCD, and gluon condensates,
respectively. In this work, we only keep the two-gluon condensate in
consideration for the heavy quark condensates are zero  as suggested
in literature \cite{Reinders:1984sr}.

Due to the quark-hadron duality, we differentiate
Eq.(\ref{R0}) with respect to $\frac{1}{M_B^2}$, then eliminate the
pole residues $\lambda_t$, and obtain the resultant sum rule for the
mass spectra of the H-H diquark states:
\begin{eqnarray}\label{massfunction}
M_{t}=\sqrt{-\frac{R_t^1}{R_t^0}}\label{mass}\;,
\end{eqnarray}
with
\begin{eqnarray}
R_t^0&=&\frac{1}{\pi}\int^{s_t^0}_{(m_{Q_1}+m_{Q_2})^2}ds
\rho_t^{\text{pert}}(s)e^{-s/M_B^2} + \hat{\bf
B}[\Pi_t^{\text{cond,4}}(q^2)]\;,
\label{R0}\\
R_t^1&=&\frac{\partial}{\partial{M_B^{-2}}}{R_t^0}\;.
\end{eqnarray}
Here, $M_B$ is the Borel parameter and $s_0$ is the threshold cutoff
introduced to remove the contribution of the higher excited and
continuum states \cite{P.Col,D.S.Du}.

For the H-H diquark states, the detailed expressions of  $R_t^0$ are
collected in the appendix.

\section{Numerical Analysis}
\subsection{The masses of the ground diquark states with only heavy flavors.}
The numerical parameters used as inputs in this work are taken as
\cite{Shifman,Wang:2010sh, Reinders:1984sr,Ioffe2005}
\begin{eqnarray}
\begin{aligned}
&\langle\bar{q}q\rangle=-(0.24\pm0.01\text{GeV})^3\;,
& &\langle\bar{s}s\rangle=(0.08\pm0.02)\langle\bar{q}q\rangle\;,\\
&\langle\bar{q}g_s\sigma G q\rangle=m_0^2\langle\bar{q}q\rangle\;,
& &\langle\bar{s}g_s\sigma G s\rangle=m_0^2\langle\bar{s}s\rangle\;,\\
&m_0^2=(0.8\pm0.2)\text{GeV}^2\;, & &\langle\frac{\alpha_s}{\pi}
G^2\rangle= 0.012\text{GeV}^4\;,\\
&m_u\simeq m_d=0.005\text{GeV}\;, & &m_s=(0.14\pm0.01)\text{GeV}\;,\\
&m_c=(1.35\pm0.10)\text{GeV}\;, & &m_b=(4.7\pm0.1)\text{GeV}\;.
\end{aligned}
\end{eqnarray}
where the energy scale is $\mu=1$ GeV.

It is crucial to determine the proper threshold $s^0$ and Borel
parameter $M_B^2$ for obtaining physical spectra. For justifying if
the choice is suitable, there are two criteria. First, the
perturbative contribution should be larger than the contributions
from all kinds of condensates, and another is that the pole
contribution should be larger than the continuum
contribution\cite{Shifman, Reinders:1984sr}. In our work, the error
bars are estimated by varying the Borel parameters, $s^0$ and
including the uncertainties of the input parameters as well.
\begin{table}[h]
\begin{center}
\begin{tabular}{|c|c|c|c|c|c|}
\hline\hline &mass(GeV) & $M_B^2$(GeV$^2$)& $\sqrt{s^0}$(GeV) &pole
&$\langle\frac{\alpha_sGG}{\pi}\rangle$\\
\hline $cc(1^+)$&$2.99\pm0.10$&$1.2-2.5$&$3.3\pm0.1$&$(56-88)\%$
&$(4-8)\%$\\
\hline $bc(0^+)$&$6.30\pm0.09$&$2.2-5.0$&$6.6\pm0.1$
&$(85-99)\%$ &$(5-14)\%$\\
\hline $bc(1^+)$&$6.36\pm0.08$&$3.0-6.0$&$6.7\pm0.1$&$(53-85)\%$
&$(3-5)\%$\\
\hline $bb(1^+)$&$9.76\pm0.08$&$6.0-14.0$&$10.1\pm0.1$&$(41-79)\%$
&$(0.04-0.23)\%$\\
\hline \hline
\end{tabular}
\end{center}
\caption{For the H-H diquark states, we show the masses, the
preferred Borel parameters $M_B^2$, the threshold parameters $s^0$,
the contribution from the pole term to the spectral density and the
contribution from
$\langle\frac{\alpha_s}{\pi}G^2\rangle$.}\label{hhmasses}
\end{table}

\begin{figure}[t]
\centering
\includegraphics[width=7cm]{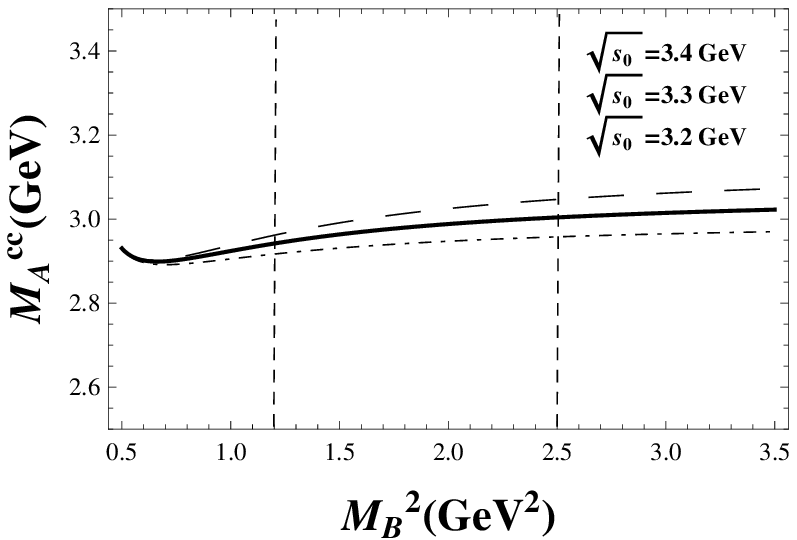}
\includegraphics[width=7cm]{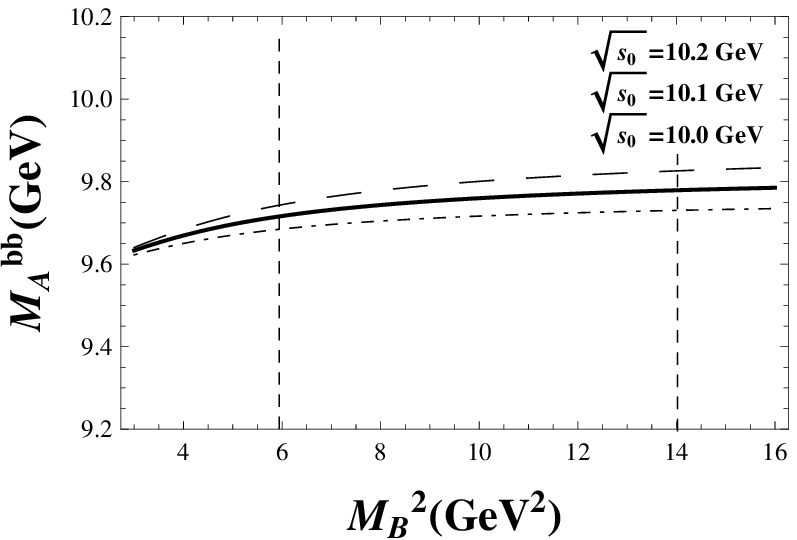}
\includegraphics[width=7cm]{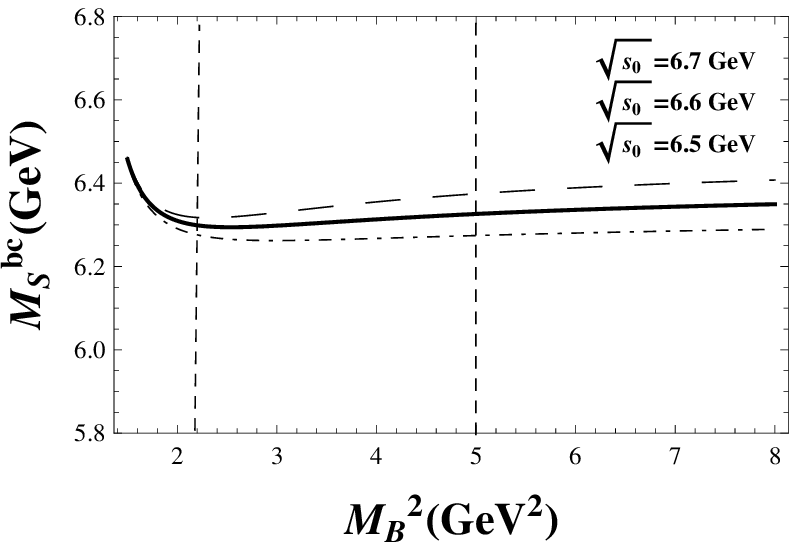}
\includegraphics[width=7cm]{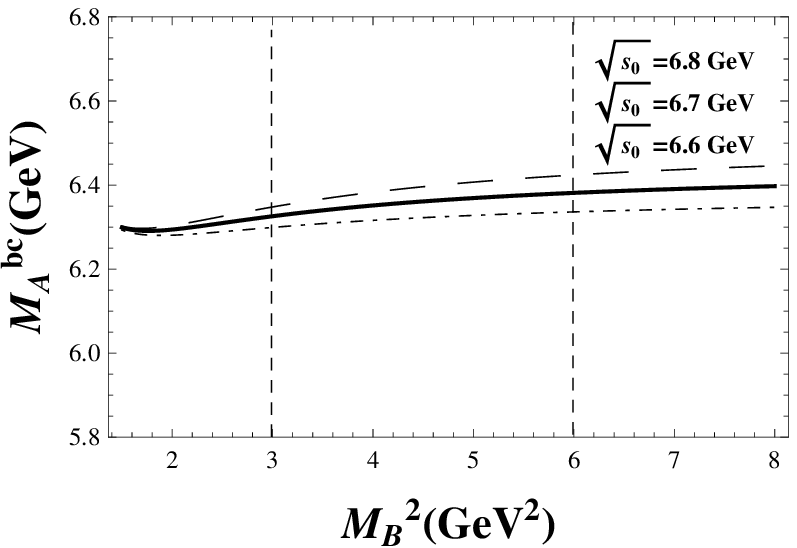}
\caption{Dependence of $M_A^{cc}$, $M_A^{bb}$, $M_S^{bc}$ and
$M_A^{bc}$ on the Borel parameter $M_B^2$. We deliberately put two
vertical lines denoting the chosen Borel window.}
\end{figure}

The spectra of the L-L  and H-L diquarks have been studied in the
previous literature\cite{Dosch:1988huJamin:1989hh,Wang:2010sh}. Our
input parameters are the same as that used in
Ref.\cite{Wang:2010sh}. Numerically we have rechecked the results of
the previous works for spectra of L-L and H-L diquarks  and find our
results are consistent with them, thus we simply present the
relevant results for the L-L and H-L diquarks in the following
table.  The analytical formulations for the H-H diquarks are shown
in the appendix and that for L-L and H-L can be found in the
relevant references which we list in the bibliography of the work.

\subsection{Discussion on the stability of the ground diquark states.}
In Table\ref{hhmasses}, we present the masses and the threshold
$\sqrt{s_0}$ of the H-H scalar and axial-vector diquark ground
states. Then, altogether with that given in literature, we collect
the results for all these three types of diquark in the following
table. Here we define the energy gap as $\Delta E=\sqrt{s_0}-M_d$,
where $M_d$ is the mass of the corresponding diquark state.

\begin{table}[h]
\begin{center}
\begin{tabular}{|c|c|c|c|c|}
\hline\hline &mass(GeV) &$M_B^2$(GeV$^2$) &$\sqrt{s^0}$(GeV)
&$\Delta E$(GeV)  \\
\hline $sq(0^+)$  &$0.55\pm0.03$ &1.0 &0.95 &0.40   \\
\hline $qq(1^+)$  &$0.34\pm0.04$ &4.0 &0.85 &0.51   \\
\hline $sq(1^+)$  &$0.42\pm0.03$ &4.0 &0.95 &0.53   \\
\hline $ss(1^+)$  &$0.50\pm0.05$ &4.0 &1.05 &0.55   \\
\hline \hline
\end{tabular}
\end{center}\caption{The diquark masses, the preferred Borel parameters $M_B$, the
preferred threshold parameters $s^0$ obtained by choosing reasonable
plateaus, and the energy gap $\Delta$E for each possible L-L diquark
state. Here the revelent results are derived from the corresponding
formulae in Ref.\cite{Dosch:1988huJamin:1989hh} with our input
parameters.}\label{ll}
\end{table}

\begin{table}[h]
\begin{center}
\begin{tabular}{|c|c|c|c|c|}
\hline\hline &mass(GeV) &$M_B^2$(GeV$^2$) &$\sqrt{s^0}$(GeV)&$\Delta
E$(GeV) \\
\hline $cq(0^+)$  &1.77 &1.50 &2.19 &0.42 \\
\hline $cq(1^+)$  &1.76 &1.60 &2.19 &0.43 \\
\hline $cs(0^+)$  &1.84 &1.55 &2.24 &0.40 \\
\hline $cs(1^+)$  &1.84 &1.65 &2.24 &0.40 \\
\hline $bq(0^+)$  &5.14 &3.85 &5.48 &0.34 \\
\hline $bq(1^+)$  &5.13 &3.95 &5.48 &0.35 \\
\hline $bs(0^+)$  &5.20 &3.95 &5.57 &0.37 \\
\hline $bs(1^+)$  &5.22 &4.10 &5.57 &0.35 \\
\hline \hline
\end{tabular}
\end{center}
\caption{The diquark masses, the preferred Borel parameters $M_B$,
the preferred threshold parameters $s^0$ obtained by choosing
reasonable plateaus for the H-L diquarks. These revelent results
have already been obtained in Ref.\cite{Wang:2010sh}.}\label{hl}
\end{table}

\begin{table}[h]
\begin{center}
\begin{tabular}{|c|c|c|c|c|c|c|}
\hline\hline &mass(GeV) &$M_B^2$(GeV$^2$) &$\sqrt{s^0}$(GeV)&$\Delta
E$(GeV) \\
\hline $bc(0^+)$  &6.30 &3.6  &6.6 &0.30 \\
\hline $cc(1^+)$  &2.99 &1.9  &3.3 &0.31 \\
\hline $bc(1^+)$  &6.36 &4.5  &6.7 &0.34 \\
\hline $bb(1^+)$  &9.76 &10.0 &10.1 &0.34 \\
\hline \hline
\end{tabular}
\end{center}\caption{The H-H diquark masses, the preferred Borel parameters $M_B$, the
preferred threshold parameters $s^0$ obtained by choosing reasonable
plateaus.}\label{hh}.
\end{table}

As aforementioned, we use energy gap $\Delta$E to embody the
stability of diquarks. Namely, according to our general knowledge on
quantum mechanics, the continuous spectra would correspond to a
dissolved state where the constituents of the supposed-bound state
are set free. Thus we choose the gap between the mass of the ground
state and the corresponding threshold $\sqrt{s_0}$ as the criterion
of stability of the diquark. In Tables(\ref{ll}-\ref{hh}), we show
the energy gap $\Delta E$ for the three types of diquarks.
Concretely, the energy gap is between 0.40 GeV and 0.55 GeV for L-L
diquarks, between 0.34 GeV and 0.43 GeV for H-L diquarks, and
between 0.30 GeV and 0.34 GeV for H-H diqaurks. One can observe that
the sequence of the energy gaps for these three types of diquarks is
:$\Delta E_{L-L}>\Delta E_{H-L}>\Delta E_{H-H}$. If larger energy
gap implies more stable structure, the L-L diquarks should be the
stablest one and the H-H diquarks is the most instable structure,
however, this definitely contradicts to our intuition. Let us
discuss this issue in next section.

\section{Summary and Conclusions}
In this work, we try to study the stability of scalar and axial
vector diquarks of three types L-L, H-L and H-H in terms of the QCD
sum rules. For the purpose, we first calculate the mass spectra of
the scalar $(0^+)$ and axial-vector $(1^+)$ H-H diquarks. Together
with the results given in the previous works about the L-L diquarks
\cite{Dosch:1988huJamin:1989hh} and the H-L
diquarks\cite{Wang:2010sh}, we can compare the corresponding $\Delta
E$ which is supposed to be a reference criterion for the stability
of the diquark sub-structure.

However, the numerical results obtained in terms of the QCD sum
rules determine the sequence $\Delta E_{L-L}>\Delta E_{H-L}>\Delta
E_{H-H}$. Even though actually the values of $\Delta E_{L-L},\;
\Delta E_{H-L}$ and $\Delta E_{L-L}$ are not far apart, this
sequence obviously conflicts to our common sense where the H-H
diquark is the most stable structure.

Because the computation with the QCD sum rules possess over 20\%
errors and the determined $\Delta E$ values do not differ much (less
than 20\%), we cannot decide that the H-H diquark is the most
instable one.

Thus, we would be tempted to conclude that due to the large errors
brought up in the computations with the QCD sum rules, it is not
suitable to determine the stability in terms of the QCD sum rules,
and we have to turn to other theoretical approaches to investigate
this important issue which would be the goal of our next work.

\vspace{.7cm} {\bf Acknowledgments} \vspace{.3cm}

When we complete the work, we notice that a new paper about the
spectra of the L-L diquark appears at the ArXiv as hep-ph 1112.5910.
Even though the concrete numbers are slightly different from that
given in Ref.[15], the general trend is similar and our qualitative
conclusion does not change at all. This work was supported in part
by the National Natural Science Foundation of China(NSFC) under
contract N0.11075079.


\appendix{\bf\Large Appendix}

For the H-H diquark, our analytical expressions are shown as
follows:
\begin{eqnarray}
\rho_S(s)&=&-\frac{3}{4\pi
s}\big(m_{Q_1}^2+2m_{Q_1}m_{Q_2}+m_{Q_2}^2-s\big)\sqrt{(m_{Q_1}^2-m_{Q_2}^2+s)^2
-4m_{Q_1}^2s}\;,
\end{eqnarray}

\begin{eqnarray}
\rho_A(s)&=&\frac{3}{2\pi
s}\big(m_{Q_1}^2-4m_{Q_1}m_{Q_2}+m_{Q_2}^2\big)\sqrt{(m_{Q_1}^2-m_{Q_2}^2+s)^2-4m_{Q_1}^2
s}\;,
\end{eqnarray}

\begin{subequations}
\begin{eqnarray}
G_S^1(M_B^2)&=&\langle \alpha_s G^2 \rangle\int_0^1 dx e^{-\frac{
\frac{m_{Q_1}^2} {x}+\frac{m_{Q_2}^2}{1-x}}{M_B^2}}\bigg\{\frac{(4-9
x) (1-x)}{16 \pi }+\frac{1}{32 \pi  x^3 M_B^2}\Big[2 (1-x)
(9 x^2\nonumber\\
&&-13 x+4) x^3 m_{Q_1}^2+18 x m_{Q_1} m_{Q_2}+2 (x-1) (9 x-4) x^4
m_{Q_2}^2\Big]+ \frac{1}{32 \pi  x^3 M_B^4}\nonumber\\
&&\times\Big[(1-x)(-9 x^5+22 x^4-17 x^3+4 x^2-7 x+7) x m_{Q_1}^4
+(6 x^3-6 x^2+3 x\nonumber\\
&&-2) m_{Q_1}^3 m_{Q_2}+(1-x) \left(18 x^4-26 x^3+8 x^2+7\right) x^2
m_{Q_1}^2 m_{Q_2}^2-6 x^3
m_{Q_1} m_{Q_2}^3\nonumber\\
&&+(4-9 x) (1-x) x^5m_{Q_2}^4\Big]+\frac{1}{32 \pi  x^3 M_B^6}
\Big[3 (1-x) (x-1)^2
\left(x^3-x^2+1\right) x^2 \nonumber\\
&&\times m_{Q_1}^6+3 \left(x^4-2x^3+x^2+x-1\right) x m_{Q_1}^5
m_{Q_2}+3 (x-1)
(3 x^4-6 x^3+3 x^2\nonumber\\
&&+2 x-2) x^3 m_{Q_1}^4 m_{Q_2}^2 -3 \left(2 x^3-2 x^2+1\right) x^2
m_{Q_1}^3
m_{Q_2}^3+(1-x) (9 x^3-9 x^2\nonumber\\
&&+3) x^4 m_{Q_1}^2 m_{Q_2}^4
+3 x^5 m_{Q_1} m_{Q_2}^5+3 (x-1) x^7 m_{Q_2}^6\Big]\bigg\}\;,\\
G_S^2(M_B^2)&=&G^S_1(M_B^2,m_{Q_1}\leftrightarrow m_{Q_2})\;,
\end{eqnarray}
\end{subequations}

\begin{subequations}
\begin{eqnarray}
G_S^3(M_B^2)&=&\langle\alpha_sG^2\rangle\int_0^1 dx
e^{-\frac{\frac{m_{Q_1}^2}{x}+\frac{m_{Q_2}^2}{1-x}}{M_B^2}}
\Big\{-\frac{9 \left(2 x^2-2 x-1\right)}{64 \pi }-\frac{1}{64 \pi
(x-1)^3 x^3 M_B^2}\nonumber\\
&&\times\Big[3 (3 x^2 \left(2 x^2-2 x-1\right) (x-1)^3
m_{Q_1}^2+8 x^2 (x-1)^2 m_{Q_1} m_{Q_2}\nonumber\\
&&-3 x^3 \left(2 x^2-2 x-1\right) (x-1)^2
m_{Q_2}^2)\Big]-\frac{1}{64 \pi (x-1)^3 x^3 M_B^4}\nonumber\\
&&\times\Big[3 (3 (x-1)^2 x^4 m_{Q_2}^4+3 (x-1)^4 x^2
m_{Q_1}^4+(x-1) x^2 (-6 x^3+12 x^2\nonumber\\
&&-7 x+1) m_{Q_2}^2 m_{Q_1}^2)\Big]-\frac{1}{64 \pi (x-1)^3 x^3
M_B^6}\Big[3 (-x^5 m_{Q_2}^6+(x-1) x^3 \nonumber\\
&&(3 x-2) m_{Q_2}^4 m_{Q_1}^2+(x-1)^5 m_{Q_1}^6+(1-3 x) (x-1)^3 x
m_{Q_2}^2 m_{Q_1}^4)\Big]\Big\}\;.
\end{eqnarray}
\end{subequations}

\begin{subequations}
\begin{eqnarray}
G_A^1(M_B^2)&=&\langle\alpha _sG^2\rangle\int_0^1 d xe^{-\frac{
\frac{m_{Q_1}^2}{x}+\frac{m_{Q_2}^2}{1-x}}{M_B^2}}\bigg\{-
\frac{(x-1)(9 x-4)}{8\pi}-\frac{1}{16 \pi  (x-1)^2 x^3 M_B^2}
\Big[-2 x^3\nonumber\\
&&\times(9 x-4) (x-1)^2 m_{Q_2}^2+2 x^2 (9 x-4) (x-1)^3
m_{Q_1}^2+36x (x-1)^2 m_{Q_1} m_{Q_2}\Big]\nonumber\\
&&-\frac{1}{16 \pi  (x-1)^2 x^3 M_B^4} \Big[(x-1) x^3 (9 x-4)
m_{Q_2}^4+(x-1)^3 \left(9 x^2-4 x-7\right) m_{Q_1}^4\nonumber\\
&&-(x-1)^2 x \left(18 x^2-8 x-7\right) m_{Q_2}^2 m_{Q_1}^2+12 (x-1)
x^2 m_{Q_2}^3 m_{Q_1}-2 (x-1)^2\nonumber\\
&&\times (3 x+2) m_{Q_2} m_{Q_1}^3\Big]-\frac{1}{16 \pi  (x-1)^2 x^3
M_B^6}\Big[-3 x^4 m_{Q_2}^6+6 x^3
m_{Q_2}^5 m_{Q_1}\nonumber\\
&&+(x-1) x^2 (9 x-3) m_{Q_2}^4 m_{Q_1}^2+3 (x-1)^4 m_{Q_1}^6+6
(x-1)^3 m_{Q_2} m_{Q_1}^5\nonumber\\
&&-3 (x-1)^2 x (3 x-2) m_{Q_2}^2 m_{Q_1}^4-6 (x-1) x (2
x-1) m_{Q_2}^3 m_{Q_1}^3\Big]\bigg\}\;,\\
G_A^2(M_B^2)&=&G^A_1(M_B^2,m_{Q_1}\leftrightarrow m_{Q_2})\;,
\end{eqnarray}
\end{subequations}

\begin{subequations}
\begin{eqnarray}
G_A^3(M_B^2)&=&\langle\alpha_sG^2 \rangle\int_0^1 dx
e^{-\frac{\frac{m_{Q_1}^2}{x}+\frac{m_{Q_2}^2}{1-x}}{M_B^2}}
\bigg\{-\frac{3 \left(6 x^2-6 x+5\right)}{32 \pi }-\frac{1}{32 \pi
(x-1)^3 x^3 M_B^2}\nonumber\\
&&\times\Big[3 \left((x-1)^3 x^2 \left(6 x^2-6 x+5\right)
m_{Q_1}^2-(x-1)^2 x^3 \left(6 x^2-6 x+5\right)
m_{Q_2}^2\right)\Big]\nonumber\\
&&-\frac{1}{32 \pi (x-1)^3 x^3 M_B^4}\Big[3 (3 (x-1)^4 x^2
m_{Q_1}^4+(x-1) x^2(-6 x^3+12 x^2-7 x\nonumber\\
&&+1) m_{Q_2}^2 m_{Q_1}^2+x^3 \left(3 x^3-6 x^2+5 x-2\right)
m_{Q_2}^4)\Big]-\frac{1}{32 \pi  (x-1)^3 x^3 M_B^6}\nonumber\\
&&\times\Big[3 \left((2-x) x^4 m_{Q_2}^6+(x-1) x^2 (3 x^2-6
x+2\right) m_{Q_2}^4 m_{Q_1}^2+(x-1)^5 m_{Q_1}^6\nonumber\\
&&-3 (x-1)^4 x m_{Q_2}^2 m_{Q_1}^4)\Big]\bigg\}\;.
\end{eqnarray}
\end{subequations}

\end{document}